\documentstyle [12pt]{article}
\evensidemargin\oddsidemargin
\begin{document}
\count0 = 1
 \title{\small{  QUANTUM MEASUREMENT PROBLEM, DECOHERENCE,
 AND QUANTUM SYSTEMS  SELFDESCRIPTION\\  }}
\small\author{S.N.Mayburov 
 \thanks{E-mail ~~ mayburov@sci.lpi.msk.su  ~~
 }\\
Lebedev Inst. of Physics\\
Leninsky Prospect 53, Moscow, Russia, 117924}
\date {}
\maketitle
\begin{abstract}
Quantum Measurements regarded in Systems Selfdescription
framework for  measuring system (MS) consist of measured state S
environment E and  observer $O$ processing input S signal. 
 $O$ regarded as quantum object which interaction with S,E
  obeys to Schrodinger equation (SE) and from it
 and Breuer selfdescription formalism
 S information for $O$ reconstructed. In particular
 S state collapse  obtained  if $O$ selfdescription state
has the dual structure $L_T=\cal H \bigotimes L_V$
where $\cal H$ is  Hilbert space of MS states $\Psi_{MS}$. $\cal L_V$
is the set with elements
$V^O=|O_j\rangle \langle O_j|$ describing random 'pointer'
outcomes $O_j$  observed by $O$ in the individual events.
The 'preferred' basis $|O_j\rangle$ defined by 
$O$ state decoherence via $O$ - E interactions.
Zurek's Existential Interpretation discussed in 
selfmeasurement framework.

\end{abstract}
\vspace{15mm}

\small {  Talk given on 'Quantum Structures' conference\\ }
\small {   Cesena,  April 2001, To appear in proceedings}
\vspace{10mm}
\section  {Introduction}

Quantum Measurement theory \cite {Busch} is now well established branch
of Quantum Mechanics (QM) with many important application mainly in Quantum
Information field \cite {Gui,Pen}. Yet its foundations still actively disputed
and most long and hot discussion concerned with
  the state  collapse or the objectification problem.
This Quantum Measurement problem, discussed here seems to be
still unresolved despite the multitude of the proposed
solutions ( for the review see $\cite  {Busch}$).
 In this paper  we  study dynamics of 
the quantum information transfer  in the measurement process
and resulting from it information restrictions.
Really,  any measurement quantum or classical is Observer
  information acquisition  about studied system S via direct or indirect
interaction with it. In classical case this interaction can be done
very small and neglected in the calculations,
 but in QM its influence can be
  quite important for  obtained measurement outcome \cite {Busch}.

 Under 
observer we mean information gaining and utilizing system 
(IGUS) of arbitrary structure  $\cite {Gui}$.
 It can be  both human brain or some automatic device 
processing the information , but in all cases it's
the system with some internal degrees of freedom (DF) excited during
 the information acquisition.
The computer information processing or perception  by human brain
supposedly corresponds to  the physical objects evolution
 which on microscopic level obeys to  QM laws.
Example  of it are electron pulses excited in computer circuits during
information bits memorization. Such correspondence for mental
processes isn't proved,  but there are now strong
experimental evidences that QM  successfully describes
  complex systems including biological ones \cite {Pen}.    
Basing on them we  concede that QM description is applicable
both for microscopic    and macroscopic
objects including observer $O$ (he, Bob) \cite {Wig}.
In the simple model dated back to Wigner
 $O$ state described
by  Dirack state vector
$|O\rangle$ ( or density matrix $\rho$ for other cases)
  relative to  some other quantum observer $O'$ (she; Alice).
 Class of microscopic measurement theories which 
account observer quantum effects sometimes called Relational QM
( for the review see  $\cite {Rov}$).
 Our  microscopic  model  
   of the measuring system (MS)  in general includes
the measured state (particle) S,
  detector D, environment E  and quantum
 observer $O$ which processes and stores the information.

The novel point of this approach is that $O$ must  describe
consistently also his own quantum state, which corresponds
to his impressions.
Observer selfdescription in the measurement process  (selfmeasurement)
can be  regarded in the context
of the general algebraic and logical problems of selfreference $\cite {Mitt}$.
In this framework Breuer derived the general selfmeasurement 
restrictions  for classical and quantum measurements  $\cite {Bre}$.
 Basing on his results we propose  here modification of standard
QM Hilbert space formalism  which 
  account observer selfmeasurement features consistently.
Its  main feature is the enlargement of QM states set $L_T$ over
standard Hilbert space $\cal H$, so that $L_T=\cal{H}$ $ \otimes L_R$
,where $L_R$  is linear space,
which elements describes $O$ information acquired in the measurement in 
individual events. 
 This modification conserves  
standard Schrodinger quantum dynamics, but permit to obtain
the subjective state collapse in S measurement.


Here it's necessary to make some  comments on our model premises
and review some terminology.
In our  model we'll suppose that MS  always can
 be described completely (including Environment E if necessary)
 by some state vector $|MS\rangle$ relative to 
$O'$ or by  density matrix for mixed cases.
 MS can be closed system , like atom in the box or open pure
system  surrounded by electromagnetic vacuum or  E of other kind.
We don't assume in our work any special dynamical
 properties of $O$ internal states beyond standard QM. 
  In this paper  the brain-computer analogy used without
discussing its reliability and philosophical 
implications $\cite {Pen}$.
We must stress that throughout our paper the observer consciousness (OC)
 never referred directly.
 Rather in our model observer $O$ can be regarded as active 
reference frame (RF) which interacts with studied object S
changing $O$ internal state and thus storing information about S.
Thus S state description 'from the point of view' of the particular $O$
referred by the terms 'S state in $O$ RF' or simply 'S  state for $O$'.
 The terms 'perceptions', 'impressions' used by us to characterize
   observer subjective description
of experimental  results 
and   defined below in strictly physical  terms \cite {Bre}.

\section {Selfmeasurement  and Quantum States Restrictions}

In Von Neuman (vN) measurement  scheme S  interacts with 
elementary  quantum detector  D   which final state becomes
 entangled with S
\cite {Busch}. In this model MS chain ended on D and observer
interaction with D supposedly is unimportant.
 We regard the simple model where   $O$ has analogous to $D$ structure
and same reaction on S input signal which permit to memorize it,
as shown below. We  omit detector D in  MS chain
assuming that S directly interacts with $O$. It's possible 
because if to neglect decoherence the only  D effect is
the amplification of S signal to make it conceivable for O.
For the start we omit also $O$-E interaction - decoherence, but later
we'll account it and study its influence. 
 In practice detector D and IGUS $O$  have many internal DFs, but
 their account  doesn't change principally the results
obtained below $\cite {May3}$.  
The example of dynamical model with many DFs
gives Coleman-Hepp model described in  $\cite {Hep}$.

 Let's consider   $O'$ description of
 the measurement performed by $O$ of binary observable $\hat{Q}$ on S state : 
$$
     \psi_s=a_1|s_1\rangle+a_2|s_2\rangle
$$
, where $|s_{1,2}\rangle$ are $Q$ eigenstates with values $q_{1,2}$.
In our model
$O$ has single effective DF  and its states space $\cal H$
contains at least three orthogonal states $|O_i\rangle$
 which are  the eigenstates of $Q_O$ 'internal pointer' observable.
Initial  $O$ state  is  $|O_0\rangle$
and  MS initial state is :
\begin {equation} 
     \Psi^{in}_{MS}=(a_1|s_1\rangle+a_2|s_2\rangle)|O_o\rangle  \label {AAB}
\end {equation}
Let's assume that  S-$O$ measuring interaction starts at $t_0$
and finished effectively at some finite $t_1$. In our model
MS evolution described by Schrodinger equation (SE).
 From SE linearity 
 the final state of MS system  relative to $O'$ observer
 for suitable  S$-O$ interaction Hamiltonian $\hat{H}_I$
 will be $\cite {Busch}$ :
\begin {equation}
   \Psi_{MS}=a_1|s_1\rangle|O_1\rangle+
a_2|s_2\rangle |O_2\rangle
                                   \label {AA2}
\end {equation}
to which corresponds the density matrix ${\rho}_{MS}$, called also
statistical state. It obeys to corresponding Schrodinger-Liouville
equation (SLE).
Thus  $|O_{1,2}\rangle$ are $O$ states induced by
   the measurement of
eigenstates  $|s_{1,2}\rangle$.
In vN theory the corresponding final S,D state is :
$\Psi_{S,D}=\sum a_i|S_i\rangle |D_i\rangle$.

 All this states including $|O_i\rangle$ belongs to MS Hilbert space
$\cal H'$ defined in $O'$ RF and  formally
Hilbert space $\cal H$ in $O$ RF
can be   obtained performing  $\cal H'$ unitary transformation $\hat{U}'$
 to $O$ c.m.s..  In  our case when we regard  only internal
or RF independent states 
 ${U}'=I$  can be taken and thus  $\Psi=\Psi^O$ for 
arbitrary states in $O'$ and $O$ RF 
correspondingly.
For $\rho$  defined on  $\cal H'$ their set denoted $L_q$;
 $O$ states $|O_i\rangle \in \cal H_O$ Hilbert subspace of $\cal H$;
 $\rho_O$ subset on $\cal H_O$ is $L_O\in L_q$.

Thus QM predicts  at time $t>t_1$  for external  $O'$ 
MS is  in the pure state  $\Psi_{MS}$ of (\ref {AA2})
 which is superposition of two states for different measurement outcomes.
MS state in $O$ RF $\Psi_{MS}^O$   obtained from
$\Psi_{MS}$ by  transformation $U'$, but as was argued
 in this case (neglecting space shift) 
 $\Psi_{MS}^O$ coincides with $\Psi_{MS}$.
 Yet we know  that experimentally   
macroscopic   $O$ observes some  random $Q_O$ value $q^O_{1,2}$
 from which he concludes
that  S final state is $|s_1\rangle$ or $|s_2\rangle$, i.e. 
S state collapses. Thus SLE  violated and can't be applied to the
measurement process.
In standard QM  with Reduction Postulate S final  state described by
the statistical ensemble of  individual final states for $O$
described by density matrix 
 of mixed state $\rho^s_m$:
\begin {equation}
 \rho^s_m= \sum_i |a_i|^2|s_i \rangle \langle s_i|
                                                              \label {AA33}
\end {equation}
to which responds in vN model random $D_i$ pointer outcomes,
described by S,D states mixture \cite {Busch}.
In our model 
we can phenomenologically ascribe to MS the corresponding mixed state :
\begin {equation}
 \rho_m= \sum_i |a_i|^2|s_i \rangle \langle s_i||O_i \rangle \langle O_i|
                                                              \label {AA3}
\end {equation}
 which principally differs from $\Psi_{MS}$.
 From $O$ 'point of view' $\Psi_{MS}$ describes superposition of two
contradictory impressions : $Q=q_1$ or $Q=q_2$ percepted simultaneously,
which Wigner claimed to be nonsense \cite {Wig}.
If observers regarded as quantum objects 
 then this  contradiction constitutes famous Wigner 
  'Friend  Paradox' for $O, O'$  $\cite {Wig}$. 
Thus MS state relative to $O$ and $O'$ looks principally
different and even contradictory, but
it's quite difficult to doubt both in correctness of
 $O'$ description of MS evolution
by Schrodinger equation and in  the state collapse experimental observations.
We attempt to reconcile this two alternative pictures 
in the united formalism which incorporate  both
 quantum   system descriptions 'from outside' by $O'$
 'from inside' by $O$ or selfdescription.

To study it, first one should introduce the relations between
IGUS  functioning and subjective information (impression).
 For realistic IGUS $|O_{1,2}\rangle$ can correspond to some
excitations of $O$ internal collective DFs
 like phonons, etc., which memorize this $Q$ information,
 but we don't consider its possible physical mechanisms here.
Concerning the relations between observer state evolution
and his information perception we use the following assumptions: 
 for any Q eigenstate $|s_i\rangle$ 
 after S measurement finished at $t>t_1$ 
and $O$ 'internal pointer'state is $|O_i\rangle$ observer $O$
 have the definite impression
that  the measurement event occurred and  input state
 is $s_i$. This calibration assumption is nontrivial and 
related to 'preferred basis' problem discussed below \cite {Elb}.   
  If S state is the superposition $\psi_s$ then we'll
suppose that its measurement also  result in appearance of some 
unspecified  at this stage $O$ impression which  will be obtained below.
Note that we don't suppose any special properties of 
biological or human systems. In our framework the simplest $O$
toy-model of information memorization is hydrogen-like atom
 for which $O_0$ is ground state
and $O_i$ are the metastable  levels excited by $s_i$, resulting so into
final S - $O$ entangled state. In this approach 'internal pointer' $O_i$
 and $O$ memory  which normally differs supposed to be the same object, but
it isn't important for our model.

Remind briefly Breuer 
theorem  results   which are valid  both for
classical and quantum measurements $\cite {Bre}$.
Any measurement of studied system $S_T$ is the 
mapping of $S_T$ states set $N_T$ on  observer states set $N_O$.  
For the situations when observer $O$ is the part of the studied system  
, $N_O$ is $N_T$ subset and 
$O$ state in this case is $S_T$ state projection on $N_O$
 called restricted  state $R_O$.  
From $N_{T}$ mapping properties the principal restrictions for
MS states recognition  by $O$ were obtained, which is the main
result of Breuer theorem.
 Namely, if for two arbitrary $S_T$ 
states $\Phi_{S}, \Phi'_{S}$ 
their restricted  states $R_O, R'_O$ coincide, then for $O$ this $S_T$ 
states are indistinguishable.
 The origin of this effect is easy to 
understand qualitatively : $O$ has less number
 of DFs than $S_T$ and so can't describe completely $S_T$ state.
 QM introduces additional features connected with observables
noncommutativity and nonlocality, some ofthem  will be  regarded
 below, but aren't important here.
For quantum measurements as $O$  restricted state
can be chosen the partial trace of MS state  (\ref {AA2}) :
\begin {equation} 
   R_O=Tr_s  {\rho}_{MS}=\sum |a_i|^2|O_i\rangle\langle O_i|
      \label {AA4}
\end {equation}
$R_O$ is in fact $\rho_{MS}$ projection into $\cal H_O$ defined
in $O'$ RF and all $R_O$ belong to $\rho_O$  set $L_O$. Note that
in quantum case Breuer theorem doesn't permit to define  restricted states 
directly; due to it
 such $R_O$ form is only the phenomenological choice and as argued below
 isn't unique possible solution.
 $R_O$ can be interpreted as $O$ subjective state which
describe his  perception of MS state after measurement,

 $R_O$  components weights are defined in $O$ basis, which
will be used in the following calculations :
\begin {equation}
w_j(t)=Tr (\hat{P}^O_j R_O)   \label {BB00}
\end {equation}
 where $\hat{P}^O_j$ is $O_j$ projection operator, and $w_j(t_1)=|a_j|^2$
for MS final state.

Note that  for MS  mixed state $\rho_{m}$
of (\ref {AA3}) the  corresponding restricted state is the same $R^m_O=R_O$.
This equality doesn't mean automatically
 collapse of MS pure state $\Psi_{MS}$,
 because as Breuer argues the collapse presence
   must be  verified by special procedure  applied to individual events.
 For this purpose it's important to
define the quantum state in the individual events; for pure MS it simply
coincides with $\Psi_{MS}$.
  Note  hence that for  incoming S mixture (\ref {AA3})
  MS individual state objectively exists  in each event $n$, but differs
from event to event  and can be described as :
$$
\rho^m (n)=|O_l\rangle \langle O_l|| s_l\rangle\langle s_l|
$$
for random $l(n)$ with probabilistic distribution $P_l=w_l(t_1)$\cite {Bre}.
This individual state can be initially unknown for $O$, but
exists objectively.
$\rho^m (n)$  differs from
statistical state (\ref{AA3}) and
its restricted state is $R^m_O(n)=R_l=|O_l\rangle \langle O_l|$  also differs 
 from $R_O$ of (\ref {AA4}). Due to it 
 the main condition of Breuer  Theorem violated,
 and this theorem isn't applicable for this situation.
Consequently $O$ can differentiate pure/mixed states 'from inside'
in the individual events \cite {Bre}.
 It means that  Breuer selfmeasurement formalism doesn't results in 
 the collapse appearance in standard QM
  even with inclusion of observer  in the
measurement model \cite {Bre}. In this framework
Straightforward $R_O$ interpretation is that it describes
$O$ internal pointer $Q_O$ 'splitting' percepted by $O$ or $O$ impressions
superposition in Wigner terms.
But it's principally important that  in this  ansatz 
MS state description by $O$ 'from inside' 
 can be really different from
  description by $O'$ 'from outside', due to  incompleteness of
$O$ selfdescription.
 Because of this incompleteness
 $O$ can't see the difference between
 physically different  pure  states with equal $D_{ij}=a_i^*a_j+a_i a_j^*$.
This situation can be called partial state collapse.
  In the formalism reported below
we'll demonstrate that changing $O$ restricted states ansatz
 it's possible to get observable S
 collapse - i.e. after S measurement $O$ can't differ pure and mixed incoming
 S states with the same $|a_i|^2$.

It's possible  to rewrite formally MS state for individual
event $n$ for $O$
for Breuer ansatz  in dual 
form $\Phi^B(n)=|\phi_D,\phi_I \gg$, where $\phi_D=\rho_{MS}$ dynamical
 state component, and $\phi_I$ is $O$ information about  MS state
acquired in event $n$. $\phi_I$ is
equal  to $R_O$ for pure state and $\phi_I=R_O^m(n)$ for mixture.  
 Of course
in this ansatz for pure state $\phi_I$ is just $\phi_D$  projection,
 but in other
selfdescription schemes  their relation can change
and the  states duality becomes principally important.
Corresponding dual states set is $N_T=L_q\bigotimes L_V$.
It's possible to define also dual
statistical state $|\Theta^B\gg=|\eta_D,\eta_I\gg$  for ensemble, for
which $\eta_D=\rho_{MS}$ and $\eta_I=R_O$ for pure state and
$\eta_I=\sum P_i |O_i\rangle \langle O_i|$ for the mixture. Its meaning
will be discussed below in more detail.
 For this purpose we'll use
  MS interference term (IT) observable :
\begin {equation}
   B=|O_1\rangle \langle O_2||s_1\rangle \langle s_2|+j.c.
    \label {AA5}
\end {equation}
In standard QM
being measured by $O'$ it gives $\bar{B}=0$ for mixed MS state (\ref {AA3})
, but in general $\bar{B}\neq 0$  for pure MS state (\ref{AA2}).
It  evidences that  for statistical ensemble 
the observed by $O'$ effects  differentiate  pure and mixed MS states.
Note that $B$ value principally can't be measured by $O$ directly, because
$O$ performs $Q_O$ measurement and $[Q_O,B] \neq0$ $\cite {May3}$. 
We'll proceed with discussion of selfmeasurement theory, in particular
description of ensembles by  restricted statistical states
 after introducing our dual formalism.

\section {Dual Selfmeasurement Formalism}

 Breuer analysis is quite  valuable, because it shows that even
observer  inclusion into measurement chain
and selfdescription restrictions account doesn't lead to collapse appearance
in standard QM. Moreover 
it prompts how to   modify standard QM formalism
so that it can describe  state collapse consistently.
The main idea is to develop alternative selfdescription formalism
   for which MS dual state  $\Phi$ becomes compatible
with S state collapse.
  
Again as the example we regard MS system consist of S and $O$
described also by external $O'$ not interacting with MS. 
From that we'll demand that our
 QM modification  satisfy to three main operational conditions : \\
i) if S (or any other system)  don't interact with $O$
 then for $O$ this  system evolves according to Schrodinger equation
dynamics  (SD) (for example GRW theory  brokes this
condition \cite {Gir}).\\
ii) If S interacts with $O$ (measurement)  SD can be violated for $O$
, but  only in such way  that  
 for stand-by $O'$     MS evolution must be
 described by SD, as follows from  condition i).\\
iii) if input S state approximates to the classical scale.
 than for $O$ and $O'$   the classical limit i.e.
objective measurement restored which for $O,O'$ is equivalent.

If  all this conditions satisfied and $O$ percepts random events
for input pure S state
   this phenomena called  the weak (subjective)
collapse. It means that in such formalism MS  final states
 relative to $O$ and $O'$ can be nonequivalent $\cite {Ume}$.
We attempt to satisfy to both this conditions by modification
of quantum state which becomes dual. It results in
 modification of QM states set, which normally is Hilbert space $\cal H$.
Remind that $\cal {H}$
is in fact empirical set  choice advocated by fitting QM data.
 Its modifications were 
published already of which most famous is Namiki-Pascazio many Hilbert spaces
 formalism \cite {Nam}.
 Analogous   superselection formalism
is well studied in nonperturbative Field theory (QFT)
 with infinite DF number $\cite {Ume}$ and were applied
 for quantum  measurement problem $\cite {Fuk,May3}$. 


To explain our main novel idea for $O$ selfmeasurement
let's consider it first for regarded MS 
measurement. Alike in regarded example for Breuer ansatz let's write
MS state in dual form $\Phi=|\phi_D, \phi_I\gg$ for dynamical and
$O$ information components, which first will be introduced
 phenomenologically. 
The first component of our dual state $\phi_D$ is also 
equal to standard QM density
matrix  $\phi_D=\rho$ and obeys always including measurement process  
to   Schrodinger-Liouville  equation (SLE)  
 for arbitrary  Hamiltonian $\hat{H}_c$ :
\begin {equation}
    \dot\phi_D=[\phi_D,\hat {H}_c]  \label {AA8}
\end {equation}
which for pure  MS states is equivalent to Schrodinger equation (SE).
 $\phi_I$, which describes the information acquired by $O$   
     differs from
Breuer $R_O$. It supposed that after S measurement in the
individual event $n$ 
    $\phi_I=V^O$ where
$V^O=|O_j\rangle \langle O_j|$ is stochastic state
 with $j(n)$ probabilistic distribution
$P_j=|a_j|^2$ and thus such dual event-state $\Phi(n)$ 
changes from event to event.
Thus  $O$ subjective information $\phi_I$ in this
phenomenological ansatz
is relatively independent of $\rho_{MS}$ and correlated with it only
 statistically.
 Clearly for such restricted states ansatz  $O$ can't differ
the pure and mixed states with the same $|a_i|^2$.
It's important to stress 
that this formalism differs principally
from standard QM state reduction, despite that both
of them results in stochastic final states.  Their difference as shown
below  in principle can be tested experimentally.

Initial state $\phi_D=\rho(t_0)$  defined also by standard QM rules.
 Before measurement starts
 $O$ state vector is $|O_0\rangle$ (no information on S) and
 the  dual state is
 $\Phi=|\rho(t_0), V_0^O \gg $ where
 $V_0^O=|O_0\rangle \langle O_0|$ is  
initial $O$ information.
$\phi_I=V^O_i$ corresponds to $\Psi_{MS}$ branch
 $|Psi_i\rangle=|O_i\rangle|S_i\rangle$
which describes MS quantum state in $O$ RF.


Complete states set in $O$ RF for this event-states is
$N_T=L_q \bigotimes L_ V$ i.e direct product of dynamical and
subjective components subsets. Here  $L_q$ is
 density matrices $\rho$ set and  $L_V$  is the  set
of diagonal  matrices  (vectors) $\phi_I$ for which
only one component $\phi_{Ij} =1$ in each event and all others are zeroes.
 If we restrict our consideration only to
pure states as we do below then $N_T$ is equivalent
  to $\cal {H} \bigotimes \mit L_V$
 and the state vector $|\Psi\rangle$
 can be used as the  dynamical component  $\phi_D$.

Of course in this approach
the quantum  states for $O'$ (and other observers)  also has
the same dual form $\Phi'$.
 $O'$ doesn't interact with MS and due to it  MS final state for her is
$\phi'D=\rho_{MS}$ of (\ref {AA2}) and
 $\phi'_I=|O'_0\rangle \langle O'_0|$. Her information
is the same before and after S measurement by $O$, because
$O'$ doesn't get any new information during it. 
 In this theory $O'$ knows that after S measurement
$O$ acquired some definite information $O_i$ but can't know without
additional measurements what this information is.
Naturally in this formalism
$O'$ has her own subjective spaces $L'_V$ 
   and in her RF  the events states
 manifold is $N'_T=\cal H' \bigotimes \mit L'_V$ for pure states.
 From the described features it's clear that subspace
$L_V$ is principally unobservable for $O'$ (and vice versa for $L'_V,O$),
 because in this formalism only  the measurement
of $\phi_D$ component described by eq. (\ref {AA8}) 
permitted for $O'$. 
But $V^O,V'^O$ can be correlated statistically via special measurement by $O'$
 of observables of dynamical component $\phi_D$.
 For this purpose $O'$ can measure
$Q_O$ on $O$ getting the information on $V^O$ content or
some other MS observables \cite {Mayb5}.
 In general if in the Universe
altogether N observers exists then the complete states manifold
described in $O$ RF  is 
$L_T=\cal {H} \bigotimes \mit L_V \bigotimes L'_V ...\bigotimes L^N_V$
of which only first two subsets are observed by $O$ directly and
all others available only indirectly via measurements on $\cal H$ substates.

From the dual state $\Phi$ one can derive
 the dual statistical state for quantum ensembles description,
because all the necessary probabilities contained in $\phi_D$.
Due to its importance it's reasonable to define it separately :
$$
|\Theta\gg=|\eta_D,\eta_I\gg=|\rho, R_V\gg 
$$
where $R_V=\sum P_i |O_i\rangle \langle O_i|$ is 
the probabilistic mixture of $V^O_i$ states describing statistics of
$O$ impressions. Generalization of this ansatz for $O'$
and other observers is straightforward.
Complete dual states set is $N_S=L_q\bigotimes L_R$, where 
$L_R$ set of  diagonal density matrixes with  $Tr\eta_I=1$, but as noticed
above $N_S$ is equivalent to $L_q$.

For dual theory $|\Theta\gg$  is analog of quantum state $\rho$ in standard QM
which  predicts the arbitrary system $S_A$ its statistical properties
and $O$ impressions statistics in its measurements.
$|\Theta \gg$ evolution for arbitrary system  Hamiltonian
is most simply expressed by the  system of equations
for its components :
\begin {eqnarray}
\frac{\partial{\eta_D}}{\partial{t}}=[\eta_D,\hat{H}_c]    \nonumber \\  
P_j(t)=tr(\hat{P}^O_j \eta_D) \label {CC} \\
R_V=\sum| P_l |O_l \rangle \langle O_l| \nonumber
\end {eqnarray}
 If $S_A$ don't interact with $O$ (no measurement)
 then $R_V$ is time invariant and one obtains
standard QM evolution for  dynamical component $\eta_D=\rho$
-  statistical state.
Thus our dual states are important only for measurement-like  processes
with direct  system $ S_A-O$ interactions, but in such case it's  the 
 analog of regarded MS system.

 The first equation of (\ref {CC}) is SLE 
 which becomes for $|\Theta\gg$
the analog of master equation for
probabilities $P_j(t)$ describing $\phi_I$   distribution.
Due to independence of MS dynamical  component $\eta_D$
 of $\eta_I$ this MS state $\Theta$ evolution is reversible and
  no experiment performed
by $O'$ on MS wouldn't contradict to standard QM.

Note that in this theory only $\Phi$ gives complete MS state
description in the individual event, from which its future state
 can be predicted. There is no contradiction that $O$ can know
both $\phi_I$ and $\phi_D$. $\phi_D$ information can be send to $O$
by $O'$ and stored in $O$ memory cells different from $|O_0\rangle$.

The  time of $V^O_0\rightarrow V^O_j$ transition for $O$ is between
$t_0$ and $t_1$  and can't be defined in the current  formalism with larger
accuracy, but it doesn't  very important at this stage.
The most plausible assumption is that $O$ perception time $t_p$ in
MS measurement has distribution :
$$
  P_p(t)=c_p \sum \frac{\partial P_i(t)}{\partial t}\quad ; \quad i\ne0
$$ 
where $c_p$ is normalization constant. Note that this result
 is compatible with standard QM.

Now let's compare our model with state
reduction in standard QM, which also  describes  how the state vector
correlated with the changes of observer information on S after
 the measurement.
 There S,D interaction induces the abrupt and irreversible
S  state  $\psi_s$ change to  random  $\psi_j$ 
and in accordance with it detector
  pointer acquires definite position $D_j$.
 This process is claimed to be objective, 
i.e. independent of any observer.
Such S,D interaction can't be described by SLE and needs to introduce
 alternative  dynamics ,  which can violate   
quantum states evolution  linearity and reversibility.
Yet it's practically impossible to  incorporate in QM
this two contradictive dynamics consistently. In distinction 
in dual formalism  the dynamical component $\phi_D$
of  MS dual state evolves linearly and reversibly
in accordance with (\ref {AA8}). This is objective evolution
in a sense that it described equivalently
relative to any observer.
Only subjective component $\phi_I$ which
 describes $O$ subjective information about $S$ changes stochastically
after S measurement - i.e S,O interaction which makes this
 formalism consistent.

Extension of dual formalism on  mixed states is obvious and  
here it presented  only for the states interested for  measurements of the kind
(\ref {AA3}). For them from eq. (\ref {CC}) naturally follows 
$P_j=|a_j|^2$ which gives $\phi_I$ distribution.
In dual formalism the restricted MS state
 $R_l=|O_l\rangle  \langle O_l|$  where 
$l(n_1)$  in  the individual event $n_1$ defined by $w_j$. It
differs from restricted state $R_O$ in standard QM given by (\ref {AA4}),
but coincide with restriction of mixed state $\rho^m_n$
 in the individual event $n$ if $l(n_1)=l(n)$. Thus Breuer
theorem condition can be fulfilled in dual formalism as expected.
Obviously in this formalism $\bar{Q}$ coincide both for $O$ and $O'$.  

For any statistical physical theory it's necessary to construct
corresponding consistent probability ansatz, including 
Kolmogorov triple. In our dual formalism 
 probabilities $P_j(t)$ coincides with corresponding standard QM 
probabilities  $P^Q_j=Tr (\hat{P}_{Sj}|\psi_s\rangle \langle \psi_s|)$
of particular outcome $q_j$
\cite {Busch}.
Remind that in standard QM any $q_j$ corresponds to D pointer position
$D_j$ and the true events for any observer are this D counts, which 
mapped on $Q$ values axe. In our formalism $q_i$ related to 
$O$ impressions $O_i$ mapped on $Q$ axe.
 Thus standard QM probabilities ansatz can be used copiously
in dual formalism.

 Dual formalism excludes spontaneous $\phi_I$ jumps
without effective S,$O$ interactions  :
 if S and O don't interact   then for $O$
 the same  random  parameter j of $V^O$ conserved. It
follows from standard QM transition probabilities for the initial
state at $t_1$ :
$$
   P'_{ij}(t_2)=|\langle \Psi_i|U(t_2-t_1)|\Psi_{j}\rangle|^2
$$
where $\hat{U}$ is unitary evolution operator.
If for all $i \ne j$ $P'_{ij}=0$ than for $O'$ MS state is uncertain,
but there is no transitions between its components in this time. For $O$
 its arbitrary state is definite i.e. some $O_j$
and continue   to be the same as at $t_1$.

To illustrate $\Phi$ dual state reversibility
 let's consider  
$gedankenexperiment$ which can be called 
 'Undoing' the measurement. Such experiment was discussed
by Deutsch $\cite {De}$ for many worlds interpretation (MWI), 
 but we'll regard its slightly different version. Its first stage coincides
with regarded
S state (\ref {AAB}) measurement by $O$ 
 resulting in the final state (\ref {AA2}). 
This S measurement  can be undone
or reversed with the help of auxiliary devices - mirrors, etc.,
which come into action at $t>t_1$ 
and reflects $S$ back in $O$ direction and make them reinteract. 
It permits for the final state $\Psi_{MS}$ obtained at time $t_1$  at the
later time $t_2$ to be transformed backward to MS initial state $\Psi^0_{MS}$.
Thus if at $t_1<t<t_2$ $O$ has information about S state;
at $t>t_2$ it's erased and MS state again is $|\Psi_{MS}^0\rangle$.
In our dual formalism  the subjective 
event-state component $\phi_I$  describes $O$ information on S
 after measurement  and at $t>t_1$  becomes equal to some random $V^O_j$.
But after reversing  independently of $j$ it
returns to initial value $V^O_0$ , according to evolution ansatz ( \ref {CC})
 described in previous chapters.
If such description of this experiment is correct, as we can
believe because its results coincides with Schrodinger MS 
evolution in $O'$ RF  
 it follows that after $q_i$ value erased from $O$ memory
 it lost unrestorably also
for any  other possible observer. If after that $O$ would measure Q again
 obtained by $O$ new value $q_j$ will have no correlation with $q_i$.
 This consideration demonstrates 
that in dual formalism to predict future MS evolution in the individual event
 $O$ should use both $\Phi$ components.

In any realistic layout to restore state (\ref {AAB}) is practically
impossible but to get the arbitrary S-$O$ factorized state by means of
such reversing is
more simple problem and that's enough for such tests.
Despite that under  realistic conditions  the decoherence processes 
make this reversing immensely difficult it doesn't contradict to
any physical laws.

 If we consider this experiment in standard QM with reduction
from $O$ point of view  we come to quite different  conclusions. 
When memorization finished at $t_1$ in each event MS collapsed to some
 arbitrary state $|s_i\rangle|O_i\rangle$. Then at $t_2$  $O$ undergoes
the  external reversing influence, in particular it can be the
second collision with S during reversing experiment and its
state changes again and such
rescattering leads to a new state correlated with $|s_i\rangle$ :
$$
      |s_i\rangle|O_i\rangle \rightarrow |s'_i\rangle|O_0\rangle
$$
It means that $O$ memory erased and he
forgets Q value $q_i$, but if he measure S state again
he would restore the same $q_i$ value.
Its statistical state is
$$
   \rho'_m=|O_0\rangle \langle O_0| \sum |a_i|^2|s'_i\rangle \langle s'_i|
$$
 But this S final state differs from
MS state (\ref {AAB}) predicted from MS linear evolution observed by  $O'$ and
in principle this difference can be tested on S state  
without $O$ measurement.

Of course one should remember that  existing for finite time intermediate
$O$ states are in fact virtual states and differ from really stable 
states  used here, but for macroscopic time
 intervals this difference becomes
very small and probably can be neglected.
In any realistic layout to restore state (\ref {AAB}) is practically
impossible but to get the arbitrary S-$O$ factorized state by means of
such reversing is
more simple problem and that's enough for such tests.
Despite that under  realistic conditions  the decoherence processes 
make this reversing immensely difficult it doesn't contradict to
any physical laws.

The analogy of 'undoing' with  quantum eraser experiment is straightforward :
there the photons polarization carry the information 
which can be erased and so change the system state $\cite {Scu}$.   
The analogous experiment with information memorization by 
some massive objects like molecules will be important test of
collapse models.
Note that observer $O'$ can perform on $O$ and S also the direct measurement
of interference terms for (\ref {AA2}) without reversing MS state.
Such experiment regarded for Coleman-Hepp model in  $\cite {May3}$
doesn't introduces any new features in comparison with 'Undoing'
and so we don't discuss it here.

\section {Decoherence and Existential Interpretation}

The preferred basis (PB) problem importance
is acknowledged in quantum measurement theory
\cite {Busch}. In its essence, $\Psi_{MS}$ decomposition on $O$,S states
in general isn't unique and so any theory must explain why namely
 $|O_i\rangle$ states appears in final mixture $\rho^m$. 
In our model PB acquires additional aspects being related
to $O$ information recognition. In the previous chapters
we made calibration assumption that $|O_i\rangle$ state percepted as $O_i$
value. But it's not clear why namely such states responds to it
and not some other  $|O^C_j\rangle$  - eigenstates of some $Q^C_O$,
belonging to another orthogonal basis. For example, it can be
$|O_{\pm}\rangle= \frac{|O_1\rangle \pm |O_2\rangle}{\sqrt{2}}$ 
for binary subspace.
 
Yet the situation changes principally, if to account decoherence - i.e.
$O$ interaction with environment E. 
It's widely accepted now that decoherence effects are very important
in measurement dynamics, and here some its features essential
for us reminded $\cite {Zur,Gui}$.  In the simplest 
decoherence model E consist of $N$ two-level systems (atoms) independently 
interacting with $O$ with $H_{OE}$ Hamiltonian, which for arbitrary
E states $|E^0\rangle$  at large $t$ gives:
 $|O^E_i\rangle |E^0\rangle  \rightarrow|O^E_i\rangle |E^0_i\rangle$
, where $|O^E_i\rangle$ belongs to orthogonal basis $O^E$
of $O$  states.  $|E^0_i\rangle$ are E  states which aren't necessarily
 orthogonal. Tuning specially measurement
Hamiltonian $H_{I}$ one can make two basises equal:
  $|O^E_i\rangle=|O_i\rangle$ and  only this case will be regarded here.
If in S measurement at $t<t_1$
 $O-$E interaction can be neglected than
under simple assumptions it results in final MS-E state :
\begin {equation}
    \Psi_{MS+E}=\sum a_i|s_i\rangle|O_i\rangle|E^0_i\rangle
                                             \label {DD1}
\end {equation}
It was proved that such triple decomposition is unique, even
if $|E_i^O\rangle$ aren't orthogonal \cite {Elb}.
 Thus PB problem formally resolved if decoherence accounted and
this is essential also  for our model.

 In addition  decoherence results in
important consequences for the mentioned perception basis choice.
Really the memorized states $|O^C_i\rangle$ excited by $S_i$ signals
 must be stable or at least long-living. But as follows from eq. (\ref{DD1})
 any  state $|O^c_j\rangle$ different from one of $|O_i\rangle$
in the short time would split into $|O_i\rangle$ combinations - entangled
$O$,E  states superposition. 
But  our calibration condition  demands that at least
$Q^O$ eigenstates will be conserved copiously and not transferred
to any combinations.
Thus in this model  $O$-E decoherence interaction selects
 the basis of long-living
$O$ eigenstates which supposedly
describes $O$ events perception and
memorization, i.e. $\phi_I$.
 In general the perception calibration by eigenstates is
very important both for our model and for quantum signal recognition studies.
It means that if our S signal is $Q$ eigenstate
 transformed to $Q_O$ eigenstate $|O_i\rangle$ then it's memorized
by $O$ for long time.
 MS-E entanglement
to some extent stabilize random $O_i$, because to erase 
$Q$ value  experimentalist, beside S,$O$  should also act on E. 
The analogous signal memorization model for brain neurons
was considered by Zurek \cite {Zur2} 

Decoherence influence should be accounted in $O$ selfdescription formalism.
In Breuer formalism $O$ interaction with E states accounted
analogously to  S states, so that
$R_O=Tr_{S,E}\rho$ derived taking trace both on $S$ and $E$ states.
 For dual formalism
 analogous approach permits to derive $| \Theta \gg$, thus defines 
 $\phi_I$ distribution. In the individual events $\phi_I$
correspons to $|O_i\rangle|S_i\rangle|E^O_i\rangle$ branch.
 In other aspects
decoherence doesn't change our selfmeasurement model.

Under realistic conditions the rate of E atoms interactions with 
detector D is very high and due to it in a very short time $t_d$
S,D partial state $\rho_p=Tr_E \rho$ becomes approximately equal
to mixed one, because $\rho_p$ nondiagonal elements becomes
very small. This fact induces the frequent claim
that the state collapse phenomena  can be completely explained by detector  
state decoherence. This decoherence collapse (DC) theory in its simplest form 
was proved to be incorrect $\cite {Desp}$, but more complicated
variants expect the study.
 The main argument against
is that to decide in which state the system is it's
necessary to analyse its complete not partial state. Omitting
details, for S,D,E it's always possible to construct IT operator
$\hat{B}$ analogous to \ref {AA5} which expectation value reveals
 that the system state and consiquently  D state is pure.
Dual formalism approach to collapse is close to  DC theory attitude,
in which also no additional reduction  postulate  used.
 The main difference
is  that DC theory claims the collapse is objective phenomena
 \cite {Desp}. In dual theory S state
 collapse has relational or subjective character and observed only by
observer inside decohering system, while for external $O'$ this system  
including E is in pure state.

The novel approach to the relation of state collapse and decoherence
  proposed in Zurek 'Existential interpretation' \cite {Zur2}.
In this approach IGUS or in our terms observer $O$ also regarded as
 quantum object
 included in the measurement chain and $O$ state decoherence
via interaction with E 'atoms' is quite important.
In the  regarded simple model 
 $O$ memorization of input signal S 
occurs in several binary memory cells $|m^j_{1,2}\rangle$,
which is the analog of brain neurons. 
  Alike in the regarded above case
this $O$ memory state suffers decoherence from surrounding E 'atoms'
which results in the system state analogous to (\ref {DD1}).  
Under  practical IGUS conditions the decoherence time $t_d$ 
is also quite small and for time  much larger than $t_d$
 $S,O$ partial state $\rho_p$ differs from the mixture
very little.  From that  Zurek concludes that $O$ 
percepts input pure S signal as random measurement outcomes.

Hence such theory conserves all the faults of discussed
above DC theory. Really analogously as it  was argued  for DC theory  
 for external $O'$ the system S,IGUS,E also
 is in the pure state even at $t\gg t_d$ and  IT observable
$B$ analogous to \ref {AA5}
  which proves it can be constructed.  Thus in standard QM framework
is incorrect to claim that IGUS percepts random events. But as easy to note
Zurek IGUS model doesn't differs principally
from our MS scheme. Due to it  dual formalism  can be applied  
to it without significant modifications. In its framework
  IGUS  subjective perception  described
by $\phi_I$ component of dual state which corresponds to random
outcomes for input pure S state.  Thus application of dual
 formalism for Zurek IGUS model supports eventually
Existential Interpretation hypothesys. Dual formalism 
considers effectively only single IGUS DF $|O_i\rangle$,
but any real IGUS includes many internal DFs practically unobservable for
him (brain molecules, etc.).  Account of their unobservability
 can make dual formalism and Zurek theory much closer practically.

\section {Collapse and System  Selfdescription}
%
%
%

Now we can regard dual formalism in the general selfmeasurement 
framework and compare it with Breuer selfdescription theory for
 standard QM. Here decoherence  neglected, because as was
demonstrated it doesn't introduces any new features after
$O$ basis was chosen.
 As was noticed Breuer theorem doesn't permit to derive MS restricted state
directly, and QM  only  demands that it satisfy to
selfmeasurement conditions i)-iii)
cited above. Thus there is some freedom of restricted states $R_O$ choice.
 The simplest such possibility is Breuer ansatz for $R_O$    
$R_O$ is  $\rho_{MS}$ projection into $\cal H_O$ defined
in $O'$ RF and all $R_O$ belong to $\rho_O$  set $L_O$.
 Such choice means in fact that $O$ restricted 
states are equivalent 'from inside' for $O$ and outside for $O'$.
 Remind that in QM the physical states set is spanned on physical
observables - Hermitian operators, which measurement
 permit to differ this states
 \cite {Br}. In QM the states difference in any RF
means corresponding observables  description and
 also the experimental measurement procedure.
  Yet Breuer doesn't propose it
 for $R_O$ and this is the weak point of this theory.
Really, $O'$ can measure all $O$ observables $A^O$
 and $\cal {H}_O$ ($L_O$) spanned
on them. For $O$ as RF (observer) it  is different and in the regarded
model its only observable is $Q_O$,
 and due to it  $O$ states set must differ from $L_O$.

Our stochastic ansatz $\Phi$ is also consistent with 
conditions i)-iii), which makes it compatible with standard QM
 in its realm. Moreover $\phi_I$ are $Q_O$ eigenstates and so
 $\phi_I$  subset $L_V \in L_O$ spanned on single
observable $Q_O$, which as was noticed is the only observable
for $O$.
 In accordance with it even ensemble statistics
$\eta_I$ doesn't contain any information about others 
$O$ or MS observables of MS state. Of course this are just semiqualitative
arguments which needs the additional mathematical clarification
and without it
 dual theory  still conserves the phenomenological features. But even
in this form its importance is in the fact that it describes
consistent mathematical formalism of state collapse in
the system selfdescription framwork.  

Note that $O$ is endpoint of MS measurement chain 
 which  can be regarded as singularity of some kind and MS quantum state
description by SLE for external $O'$ is regularization of this singularity
which defines its properties consistently. As shows
experience  with QFT  the singularities can results in appearance of
new regimes in particular new stochastic parameters \cite {Ume}. The
close analogy to discussed effects seems Spontaneous Symmetry Breaking
 phenomena in nonperturbative fields interactions.

The interesting interpretation problem of dual QM formalism is:  can we tell
that in given event other $\Psi_{MS}$ branches exist beside
$|O_i\rangle |S_i\rangle$ observed by $O$ ? To decide it
note that other branches existence for $O'$ can be confirmed by 
by $B$ measurement on MS and for $O$ indirectly in
discussed 'undoing' experiment. But this branches coexistence
differs from MWI where each of this branches exists in one
of many  parallel worlds. In dual theory all this branches
coexist in the same single world,
 but other branches aren't percepted by $O$, due to principal
incompleteness of $O$ selfdescription. This isn't too 
surprising, because
realistic $O$ has many internal DFs i.e. $O$ substates
 practically unobservable for him  \cite {May3}.

 Note that subjective  $\phi_I$ component of $\Phi$ 
 isn't the new degree of freedom, but   $O$
additional information about its own observable $Q_O$ and correspondingly
 on S observable $Q$ which   doesn't
  contained in state vector $\Psi_{MS}$.
 $Q$ information  for $O'$ which don't interact
with S described by $\Psi_{MS}$  and corresponds to arbitrary uncertainty
$q_{min} < Q < q_{max}$.  $\phi_I$   contains additional information
 on random $Q=q_j$ percepted by $O$ only.
 In this dual theory Schrodinger dynamics and  state
collapse coexist, by the price
 that S signal perception by $O$ occurs via this new stochastic
   mechanism. Despite we use term 'perception' in our model it doesn't 
referred to human brain specifically. We suppose that as $O$ can be
regarded any system which  can produce the stable entanglement of
its internal state and measured state S. 
It can be even  hydrogen-like atom in the simplest case for 
which $O_i$ can be different atomic levels.  Its perception
corresponds to $\phi_I$ state component  and means that in such formalism
 $O$ state differently
described  by interacting $O$ 'from inside'  and  by any other $O'$.


In the dual theory S state collapse has subjective or relative
 character and due to it strictly connected with acquired S information.
To illustrate relation between  state collapse and information
transfer we consider several  $gedanken$ experiments.
Remind that standard QM reduction postulate 
 settles that if Q acquired after measurement 
the definite value relative to 
$O$ then its objectively exists also for $O'$ or any other
observer, but can be unknown for them.
 In first experiment at the initial stage   $O$ measures $Q$ value of
 S at $t_1$ which results in
MS state (\ref {AA2}) for $O'$, but after it  $Q$ is measured again 
by observer $O'$ at $t_2>t_1$.
The interaction of $O'$ with MS results in entangled state of 
S,$O$ and $O'$  and so both observers acquire  some information 
about S state. This state vector in our formalism is:
\begin {equation}
   \Psi'_{MS}=|a_1 |s_1\rangle|O_1\rangle|O'_1\rangle+
a_2|s_2\rangle|O_2\rangle|O'_2\rangle
                                   \label {AAX}
\end {equation}   
   Analogous experiments  was discussed  frequently due to
its relation to EPR-Bohm correlations \cite {Aha2} and here we regard
 only its timing aspect. Our question is : at what time
$Q$ value becomes definite and thus S state collapse occurs for $O'$ ? 
In our formalism at $t_1<t<t_2$ observer $O$ already acquired the information
that Q value is some  $q_i$, reflected by $\phi_I=|O_i\rangle\langle O_i|$.
 In the same time Q value stays
 uncertain  for $O'$, because relative to her MS state vector is (\ref {AA2})
,and $O'+$ MS dual state for $O'$ is 
$$
     \Phi'=| \Psi_{MS}\otimes|O'_0\rangle , |O'_0\rangle \langle O'_0| \gg
$$
 When at $t>t_2$  measurement by $O'$  
 finished  Q value measured by $O'$ 
coincides with $q_i$.
To check that Q value coincides for $O'$ and $O$, $O'$ can perform 
measurement both Q and $Q_O$ which is described by (\ref {AAX})
and gives the same result as in standard QM.  
It don't contradicts to the previous assumption 
 that for $O'$ before $t_2$ $Q$ was principally uncertain.
The reason is that in between $O'$ interacts with S and it
 makes Q value definite for her. 
This measurement   demonstrates the subjective character
of collapse, which happens only after  S interaction  with
particular observer occurs.
If collapse occurs according to QM reduction postulate 
 then at $t>t_1$ MS state relative to $O'$ must be
the mixture $\rho_m$ of (\ref {AA3}). 
In our formalism at that time MS state vector relative to
$O'$  is pure state $\Psi_{MS}$ of (\ref {AA2})  which isn't $Q$
eigenstate.
 To test it experimentally $O'$ can measure $\hat{B}$ on MS at $t>t_1$
which don't commute with Q. If our theory is correct then 
$\bar{B} \ne 0 $ and thus MS state collapse doesn't occurs
at $t=t_1$.  

Remind that the state vector has two aspects : dynamical
 and  informational in which $\Psi$ is $O$ maximal information
about the  object S \cite {Busch}. Our formalism extends this aspect on
the case when $O$ measures S and can acquire more information about S
then 'stand-by' $O'$. In its framework the state collapse directly
related with $O$  information acquisition via interaction with S. 
 The same information after S measurement can be send by  $O$ to  $O'$
 in form of  some material signal, for example photons bunch.
 When $O'$ performs quantum measurement of this signal
 it result for her into   $\Psi_{MS}$ collapse
to one of final outcome eigenstates, which in our formalism
reflected by $\phi_I$ change. Thus in our theory S state collapse
directly related with information transferred to arbitrary $O$ via  
interaction with S or some intermediate system (signalling).

Relativistic analysis of EPR-Bohm pairs measurement  also indicates
subjective character of state vector and its collapse $\cite {Aha2}$.
 It was shown
that  the state vector can be defined only on space-like hypersurfaces
which are noncovariant for different observers.
This results correlate with nonequivalence of different observers
 in our  nonrelativistic formalism.
Hence we believe EPR-Bohm correlations
 deserve the detailed study in this dual framework.

\section {Discussion}

In this paper the measurement models which accounts observer (IGUS) information
processing and memorization regarded. Real IGUSes are very complicated systems 
with many DFs, but the main quantum effects 
 can be   studied with the simple models. 
The presented dual formalism demonstrates that probabilistic realization
 is generic and unavoidable for QM and without it QM supposedly can't
acquire any operational meaning. Wave-particle dualism
was always regarded as characteristic QM  feature, but in our formalism
it has straightforward  correspondence in dual states ansatz.

Breuer formalism shows that inclusion of observer as quantum object
into measurement scheme doesn't lead to collapse appearance \cite {Bre}.
To make it possible in our  approach  observers 
 are made nonequivalent
in a sense that the  physical reality description can be principally different
for each of them \cite {Rov}. This nonequivalence
reflected by the presence of subjective component $\phi_I$ available directly
only for particular $O$ and for him the subjective state collapse can be
obtained.  
Our theory indicates that to obtain it it's necessary also
 to modify the quantum states set
which makes it nonequivalent for different observers but conserves
Schrodinger evolution for arbitrary quantum system.
Here we note  that in this framework the key problem becomes  the
existence of the objective reality of the physical objects
properties  
independent of particular observer. Our results hints that no
such reality is possible and any such property has only subjective reality
relative to particular observer.

The natural question arise : does observation of random
outcomes $\phi_I=V^O_j$ means that before the measurement starts 
 S state can be characterized by some objective  'hidden parameter'
$j_S$ ? Our formalism is principally different from Hidden Parameters
theories where this stochastic parameters influence quantum state
dynamics and  so differs from SD. Due to it in our 
 model  $\phi_I$ internal parameter $j$ can be 'generated' 
during S-$O$ interaction and don't exists objectively before it starts.

As we supposed in the introduction our theory doesn't need any
addressing to to human observer consciousness (OC). Rather in this model 
$O$ is active RF which internal state excited by the interaction
with the studied object. 
This approach to the measurement problem 
has much in common with Quantum reference frames introduced by
 Aharonov  $\cite {Aha3}$.

The ideas close to our dual theory were  discussed in QM modal
interpretation, but they have there phenomenological and
philosophical formulations \cite {Busch}. Now this is the whole class
of different theories, of which the most close to us 
is  Witnessing interpretation by Kochen $\cite {Koch}$.
His theory phenomenologically supposed that for apparatus $A$
measured value $S$ in pure state always has random definite value
$S_j$, yet no physical arguments for it and no mathematical formalism differ
from standard QM  were proposed.

Historically the possible influence of observer on measurement
process was discussed first by London and Bauer \cite {Lon}. They supposed
that OC due to 'introspection
action' violates in fact Schrodinger equation for MS 
and results in state reduction. This idea was criticized in detail 
by Wigner \cite {Wig}. In distinction in our dual theory OC perception     
doesn't violate MS Schrodinger evolution from $O'$ point of view.
But measurement  subjective perception in it also performed by OC
and its results partly independent of dynamics due to
its dependence on stochastic $V^O$. This effect deserves further
discussion, but we believe that  such probabilistic behavior
is general IGUS property not related to OC only.

Dual formalism deserves comparison with  different
Many Worlds interpretations (MWI) 
variants, due to their analogy  - both are the theories without
dynamical collapse  $\cite {Busch}$. MWI is still very popular,
despite its serious consistency problems. The most close to our theory seems
 Everett+brain MWI interpretation in which eq. ($\ref {AA3}$)
describes so called observer $O$ splitting identified with state collapse
 $\cite {Whi}$.  In its framework  assumed that each
$O$ branch describes the different reality - 
separate Universe and the state collapse is
phenomenological property of human consciousness. Obviously this
approach has
some common points with our models which deserve further analysis. 
In this terms our theory can be qualified as MDI - Many Descriptions
interpretation - stressing that in it the picture of the same Universe
for different observers can differ principally.

In general all our experimental conclusions are based on human
subjective perception. Assuming the computer-brain perception analogy
in fact means that human signal perception also defined by $\bar{Q}_O$
values. Despite that this analogy looks quite reasonable we can't
give any proof of it.  In our model
in fact  the state collapse have subjective character and 
occurs initially only for single observer $O$ $\cite {Rov}$.
We present here very simple measurement theory and we don't regard
it as final solution of measurement problem. Yet from its results
we believe that it's impossible to solve it without account of 
$O$ interaction with measured system at quantum level \cite {Zur2}.


\end {document}